\documentclass[%
 reprint,
 amsmath,amssymb,
 aps,
]{revtex4-2}

\usepackage{graphicx}% Include figure files
\usepackage{dcolumn}% Align table columns on decimal point
\usepackage{bm}% bold math
\usepackage{hyperref}% add hypertext capabilities
\usepackage{upgreek}
\begin{document}

%\preprint{APS/123-QED}

\title{Recovery of topologically robust merging bound states in the continuum\\in photonic structures with broken symmetry}

\author{Huayu Bai}
\email{huayu.bai@aalto.fi}
\author{Andriy Shevchenko}%
 \email{andriy.shevchenko@aalto.fi}
 \author{Radoslaw Kolkowski}
 \email{radoslaw.kolkowski@aalto.fi}
\affiliation{%
Department of Applied Physics, Aalto University, P.O.Box 13500, Aalto FI-00076, Finland
}%

\begin{abstract}
Optical bound states in the continuum (BICs) provide a unique mechanism of light confinement that holds great potential for fundamental and applied research in optics and photonics. Of particular interest are merging BICs realized in planar periodic structures by merging accidental and symmetry-protected BICs. Topological nature of merging BICs renders their $Q$ factors exceptionally high and robust. However, the existence of accidental BICs relies on the up-down mirror symmetry of the structure. If this symmetry is broken, e.g., by a substrate, the $Q$ factor of the mode drops down. Consequently, ultrahigh-$Q$ merging BICs cannot be achieved in substrate-supported structures. Here, by studying the case of a one-dimensional periodic dielectric grating, we discover a simple method to fully compensate for the detrimental effect of breaking the up-down mirror symmetry. The method makes use of a thin layer of a high-refractive-index dielectric material on one side of the structure, allowing one to restore the diverging $Q$ factor of the accidental BIC and fully recover the merged BIC. By investigating the far-field polarization patterns of the modified gratings, we show that the integer-charge polarization vortices of the accidental BICs are restored by intersecting the momentum-space trajectories of circularly polarized half-vortices simultaneously in the upward and downward radiation directions. Our approach can enable flexible design, feasible realization, and extended applications of topologically robust BICs in various systems.

\end{abstract}

\maketitle

\section{Introduction}

Efficient confinement of light in nano- and microstructures plays an important role in many areas of optical science and technology~\cite{koenderink15}, including lasers~\cite{wang17}, nonlinear optical devices~\cite{kolkowski23}, and sensors~\cite{gallinet13}. One of the most efficient ways to trap light is to couple it to specific resonant modes known as optical bound states in the continuum (BICs)~\cite{marinica08,hsu13,hsu16,joseph21,azzam21,kang23}. By analogy with their quantum-mechanical precursors~\cite{vonneumann29,friedrich85}, BICs operate on the basis of cancelling the outgoing radiation via destructive interference. The $Q$ factors of BICs are theoretically unlimited, as long as other types of optical loss (e.g., absorption) are also eliminated~\cite{kolkowski23_}. Indeed, $Q$ factors on the order of 10$^6$ have been experimentally demonstrated with BIC excitations in dielectric photonic crystal slabs~\cite{jin19,chen22}. BICs are extensively explored in view of their potential applications in lasers~\cite{kodigala17,ha18,huang20,hwang21,mohamed22}, sensors~\cite{romano19,maksimov22}, nonlinear frequency converters~\cite{carletti18,anthur20,wang20,zograf22}, and quantum-optical devices~\cite{santiago22}.

In planar periodic structures, BICs correspond to the centers of far-field polarization vortices observed in the momentum space. These vortices are characterized by integer topological charges~\cite{zhen14,doeleman18,li19,wang20_}. The most common type of BICs are symmetry-protected BICs, which are formed at high-symmetry points in the momentum space and result from their symmetry mismatch with free-space radiation. The other type of BICs are accidental BICs that originate from coupling of two different optical modes, e.g., a guided mode and a localized Fabry-P\'{e}rot mode~\cite{hu22}. The vortex charge conservation makes accidental BICs robust against small changes of the structure parameters, which simply move the BIC to a different location in the momentum space. This allows different types of BICs to be overlapped, e.g., at the $\Gamma$-point. The resulting coalescence of several polarization vortices gives rise to the so-called merging BICs~\cite{jin19}, which can exhibit strongly increased $Q$ factors over an extended range of parameters. Such BICs have been previously investigated in photonic structures of various geometries~\cite{jin19,kang21,kang22,wan22,huang22,lee23,bulgakov23,sun24}, and their superior topological robustness made them an attractive platform for application in lasers~\cite{hwang21,cui23}, second-harmonic generation~\cite{liu23,ge23,qi23}, and sensing~\cite{liu22,barkaoui23}.

Despite their topological stability, BICs are sensitive to symmetry distortions~\cite{koshelev18,overvig20}. In the case of accidental BICs, which are the tunable ingredients of the merging BICs, breaking the up-down mirror symmetry ($\sigma_z$ symmetry) transforms the far-field vortices into pairs of ``half-vortices'', i.e., circularly polarized states with half-integer topological charges~\cite{liu19,yoda20,yin20,ye20,yuan21,zeng21,abdrabou21,wang22,liu23_}. This leads to non-zero radiative loss~\cite{zhang22,shi22,chen23}, turning accidental BICs into leaky modes (quasi-BICs) with limited $Q$ factors~\cite{sadrieva17,xu24,glowadzka21}. It has been shown that, by additionally breaking the in-plane symmetry, it is possible to restore the polarization vortex for one of the two radiation directions (either up or down)~\cite{yin20}. However, to our knowledge, none of the previous works reported the possibility to realize merging BICs in structures with broken $\sigma_z$ symmetry. Nearly all structures fabricated on a substrate lack this symmetry~\cite{xu24,glowadzka21}, whereas fully symmetric structures, e.g., free-standing or deeply buried, are difficult to fabricate and use.

\begin{figure}[hbt!]
\centering\includegraphics[width=8.6cm]{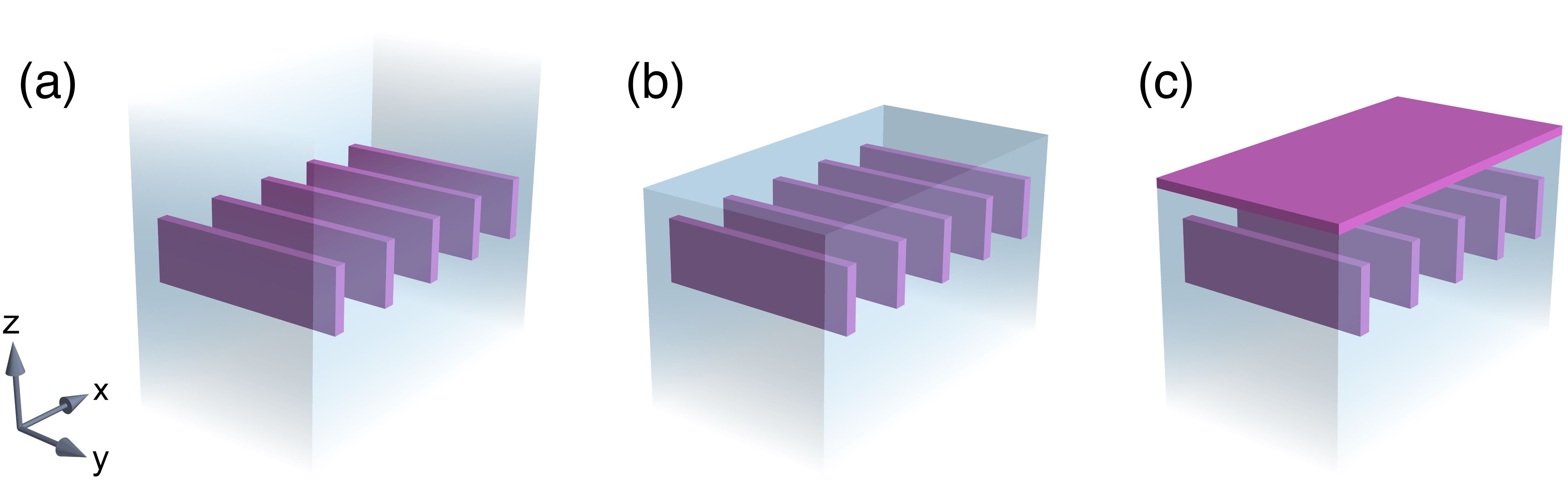}
\caption{\label{fig:1} Illustration of dielectric gratings considered in this work, with high-refractive index material (Si) shown in purple, and low-refractive index host medium (SiO$_2$) shown in pale blue. Each geometry is invariant along the $y$-axis and periodic along the $x$-axis. (a) A grating with up-down mirror symmetry provided by infinite claddings on both sides. (b) Breaking the symmetry by an interface with air on the top side. (c) Compensation for the effect of the broken symmetry by an additional layer of a high-index dielectric. }
\end{figure}

In this paper, we investigate BICs that appear as optical eigenmodes of one-dimensional (1D) periodic dielectric gratings made of a high-index material in a low-index host medium (see Fig. \ref{fig:1}). The $\sigma_z$-symmetric gratings [Fig. \ref{fig:1}(a)] are found to support pairs of off-$\Gamma$ accidental BICs (each with topological charge $+1$) that can be tuned to merge with a symmetry-protected BIC at the $\Gamma$-point (charge $-1$), giving rise to a merged BIC (charge $+1$). In practice, this can lead to increase of the $Q$ factor in its vicinity by several orders of magnitude. Breaking the symmetry by an upper interface with air [Fig. \ref{fig:1}(b)] splits each accidental BIC vortex into two half-vortices corresponding to circularly polarized radiated waves. This destroys the merged BIC and its robustness. However, we manage to fully recover both the diverging $Q$ factors of the accidental BICs and the topological robustness of the merged BIC by a simple modification of the refractive index distribution in the structure, namely, by coating the interface with a thin layer of a high-index material [Fig. \ref{fig:1}(c)]. This enables ultrahigh-$Q$ accidental and merging BICs to be realized in the absence of the up-down mirror symmetry, e.g., in conventional substrate-supported photonic structures, unlocking their potential for many applications. More broadly, our findings suggest a general approach to design systems that exhibit symmetry-protected topological properties despite breaking the essential symmetry.

\begin{figure*}[hbt!]
\centering\includegraphics[width=15.128cm]{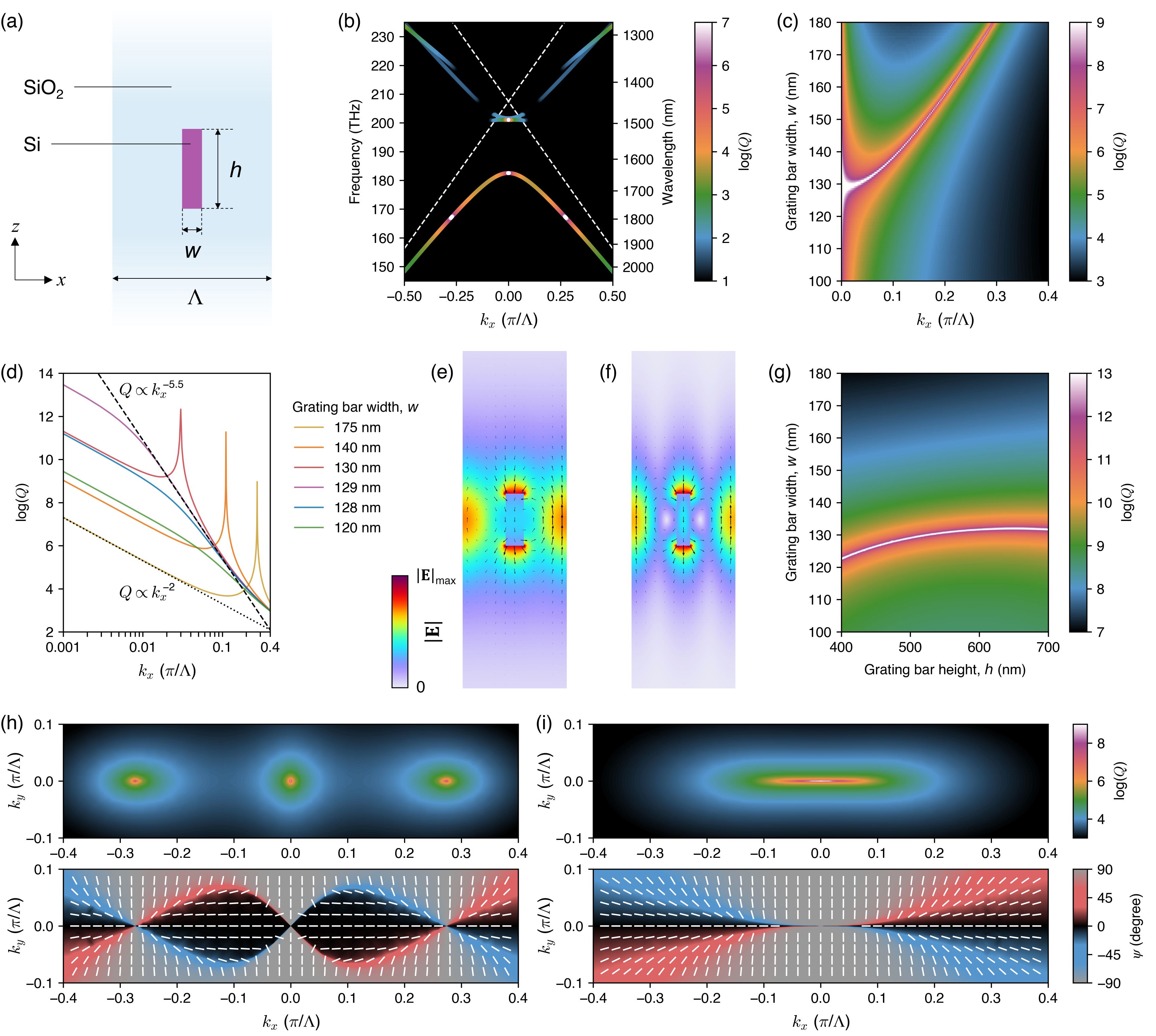}
\caption{\label{fig:2} Accidental, symmetry-protected, and merging BICs in 1D periodic dielectric gratings with up-down mirror symmetry. (a) Schematic illustration of the unit cell of the grating. The grating period $\Lambda$ is fixed at 1000 nm. (b) The band structure of the grating with $w$ = 175 nm and $h$ = 500 nm. The $Q$ factor is encoded in the color, showing the accidental and symmetry-protected BICs along the lower-frequency band. The eigenfrequencies are plotted on the black background, making the modes with $Q<10$ invisible. (c) The $Q$ factor of the lower-frequency band as a function of $k_x$ and $w$. The merging point is visible at $w$ = 129 nm. (d) Log-log plot of the $Q$ factor versus $k_x$ for several values of $w$ [horizontal cross sections through (c)]. The scaling laws of $Q\propto1/k^{5.5}$ and $Q\propto1/k^2$ are shown by the dashed and dotted black lines, respectively. (e) and (f) Electric field distributions (color for $\left|\mathbf{E}\right|$ and arrows for instantaneous $\mathbf{E}$) of the accidental and merging BICs, respectively. (g) The $Q$ factor at $k_x$ = 0.001 $\pi/\Lambda$ as a function of $h$ and $w$, showing the trajectory of the merging point. (h) and (i) Momentum-space distributions of the $Q$ factor (top plots) and the orientation of the long axis of the polarization ellipse defined by Eq.(\ref{eq:psi}) (bottom plots) before and after BIC merging, respectively.}
\end{figure*}

\section{Results and discussion}

\subsection{Accidental, symmetry-protected, and merging BICs in gratings with up-down mirror symmetry}

The existence of merging BICs in simple 1D gratings possessing $\sigma_z$ symmetry has been demonstrated in previous works~\cite{bulgakov23, bai24_}. Here, as an example of such a system, we consider a Si grating in an infinite SiO$_2$ medium. The unit cell of the grating is presented in Fig. \ref{fig:2}(a). Its geometry is determined by three parameters: the width $w$ and height $h$ of the Si bar, and the period $\Lambda$, which we fix at $\Lambda$ = 1000 nm throughout the paper. We assume that the refractive index of SiO$_2$ follows the dispersion formula by Malitson~\cite{malitson65,polyanskiy24}, while the refractive index of Si is described by the tabulated data for $T$ = 293 K by Li~\cite{li80,polyanskiy24}. We study the optical modes of this system in the infrared spectral range, in which the optical absorption loss can be neglected. Previously, we have considered a Si$_3$N$_4$ grating supporting accidental and merging BICs in the visible range~\cite{bai24_}. 

We study the photonic modes of the grating using the numerical eigenfrequency analysis with Floquet periodic boundary conditions implemented in the COMSOL Multiphysics software. As an example, Fig. \ref{fig:2}(b) shows a calculated band diagram (real part of the eigenfrequency versus the in-plane momentum $k_x$, with the $Q$ factor encoded in the color) for a grating with $w$ = 175 nm and $h$ = 500 nm. The diagram reveals several photonic bands, resulting from folding of the TE- and TM-polarized guided modes to the light cone. The higher-frequency bands appear discontinuous due to the strong radiation loss introduced by the first-order diffraction above the folded light lines (plotted as white dashed lines). The BICs are clearly visible as the $Q$ factor maxima (white points). In particular, the lower-frequency photonic band that is not affected by the first-order diffraction has one symmetry-protected BIC at the $\Gamma$-point and two accidental BICs at $k_x$ = $\pm$ 0.274 $\pi/\Lambda$. Contrary to the accidental BICs, the symmetry-protected BIC does not rely on the $\sigma_z$ symmetry, but its existence is secured by the in-plane symmetry, i.e., $C_2^zT$ (where $C_2^z$ means 180$^\circ$ rotation around the $z$-axis, and $T$ means time reversal)~\cite{zhen14}. One of the higher-frequency bands also supports a symmetry-protected BIC. However, in this paper, we focus at the BICs in the lower-frequency band only.

By keeping $h$ fixed at 500 nm and decreasing $w$, the two accidental BICs are moved towards the $\Gamma$-point along the band and merged with the symmetry-protected BIC, creating a single merged BIC. Figure \ref{fig:2}(c) presents the $Q$ factor as a function of $k_x$ and $w$ for positive $k_x$. The color map clearly reveals the trajectory of one of the accidental BICs and the merging point at $w$ = 129 nm. It is noteworthy that the range of ultrahigh $Q$ values in the $k_x$-$w$ parameter space is significantly extended around the merging point. 

Figure \ref{fig:2}(d) presents several horizontal cross sections of the plot in Fig. \ref{fig:2}(c), displaying the $Q$ factor versus $k_x$ in a log-log plot for $h$ = 500 nm and a few selected values of $w$. Regardless of the choice of parameters, the $Q$ factor of the symmetry-protected BIC diverges to infinity at $k_x$ = 0. The same is true for the accidental BICs in the case of perfectly $\sigma_z$-symmetric gratings. The finite height of their peaks in Fig. \ref{fig:2}(d) is  due to a finite step size in sampling the $k_x$-axis. 

Examples of the electric near-field distributions for the accidental and merging BICs are presented in Figs. \ref{fig:2}(e) and \ref{fig:2}(f), respectively. These plots show that the studied BICs belong to a TM photonic band, with the electric field vector lying in the $x$-$z$ plane. The accidental BIC exists at non-zero $k_x$, which leads to a phase shift across the unit cell [Fig. \ref{fig:2}(e)]. The near-field for the symmetry-protected BIC is not shown, but it is very similar to that of the merged BIC [Fig. \ref{fig:2}(f)]. 

To quantify the $Q$ factor improvement by BIC merging, one usually investigates the $Q$ factor scaling law in the vicinity of the BIC. In particular, the $Q$ factor scales as $1/k_x^2$ for the symmetry-protected BIC before merging ($w$ = 175 nm), whereas at the exact merging point ($w$ = 129 nm), the scaling law is boosted up to $1/k_x^{5.5}$, which is close to $1/k_x^6$ expected for such a merged BIC~\cite{jin19}. It has been shown that the modified scaling law of merging BICs makes them significantly more robust against random defects that break the in-plane symmetry~\cite{jin19}. Furthermore, in the case of a finite 1D grating with $N$ periods, the $Q$ factor scales as $N^3$ for a merged BIC, as opposed to $Q\propto N^2$ for a symmetry-protected BIC~\cite{bulgakov23}.

As can be seen in Fig. 2(d), the scaling law appears to be modified only over a narrow range of $k_x$. Therefore, it is more useful to investigate the magnitude of the $Q$ factor at a fixed $k_x$ close to the BIC, e.g., at $k_x$ = 0.001 $\pi/\Lambda$, for which all the curves essentially return to the $1/k_x^2$-dependence. Since the values of the $Q$ factor in this range are not realistic in view of experimental demonstration ($Q>10^7$), we only discuss the relative increase of the $Q$ factor at $k_x$ = 0.001 $\pi/\Lambda$. For example, we find that the $Q$ factor of the merged BIC at $w$ = 129 nm is improved by approximately six orders of magnitude compared to the symmetry-protected BIC far from merging ($w$ = 175 nm). Unfortunately, such an extreme $Q$ factor boost is only achievable by precise tuning of $w$. If $w$ is off by $\pm$1 nm, the $Q$ factor is increased only by four orders of magnitude. From the nanofabrication point of view, a realistic tolerance in $w$ is $\pm$ 10 nm. In such a case, the $Q$ factor is improved by two orders of magnitude, which is not as impressive as the six-order boost for a perfect merged BIC, but is still remarkable.

The $Q$-factor boost at $k_x$ = 0.001 $\pi/\Lambda$ can be used to precisely locate the merging point in the parameter space. For example, Fig. \ref{fig:2}(g) shows the $Q$ factor at $k_x$ = 0.001 $\pi/\Lambda$ as a function of $w$ and $h$, revealing a continuous trajectory of the merging condition in this parameter space. For each given $h$, the maximum $Q$-factor boost of six orders of magnitude is only achievable in a narrow range of $w$. However, due to the flattening of the trajectory at $w$ $\approx$ 650 nm, the merging point can be made almost insensitive to $h$. Assuming that $\Lambda$ is well controlled (which would be the case in most lithography-based nanofabrication techniques), precise tuning of $w$ remains the only challenge in achieving the exact BIC merging in the studied system.

Figure \ref{fig:2}(h) shows the $Q$-factor distribution across the 2D momentum space ($k_x$-$k_y$) for the grating with $w$ = 175 nm and $h$ = 500 nm (top plot), and the corresponding far-field polarization pattern (bottom plot) showing the orientation of the long axis of the polarization ellipse. The tilt angle $\psi$ of this axis with respect to the $x$-axis is determined from
\begin{equation}
\label{eq:psi}
    \tan{2\psi}=\frac{2 \operatorname{Re}\left(E_x E_y^*\right)}{\left|E_x\right|^2-\left|E_y\right|^2},
\end{equation}
where $E_x$ and $E_y$ are the Cartesian components of the electric field far from the grating. In the case of a $\sigma_z$-symmetric grating, the radiation patterns in both the upward and downward directions are identical. The patterns clearly show the polarization vortices associated with the BICs. Their topological charges can be easily determined by counting the number of full counter-clockwise rotations completed by the long axis of the polarization ellipse following a closed trajectory around a given vortex in the counter-clockwise direction. In particular, the symmetry-protected BIC is found to have a topological charge of $-$1, whereas each of the accidental BICs has a topological charge of $+$1. Figure \ref{fig:2}(i) shows that merging of the accidental BICs with the symmetry-protected BIC (by decreasing $w$) gives rise to a single polarization vortex that is strongly stretched along the $k_x$-axis, corresponding to the modified scaling law presented in Fig. \ref{fig:2}(d). The topological charge of this vortex is $+$1, which is equal to the sum of the charges of the constituent vortices, in accordance with the global charge conservation.

\subsection{Breaking the up-down mirror symmetry}

\begin{figure*}[hbt!]
\centering\includegraphics[width=15.327cm]{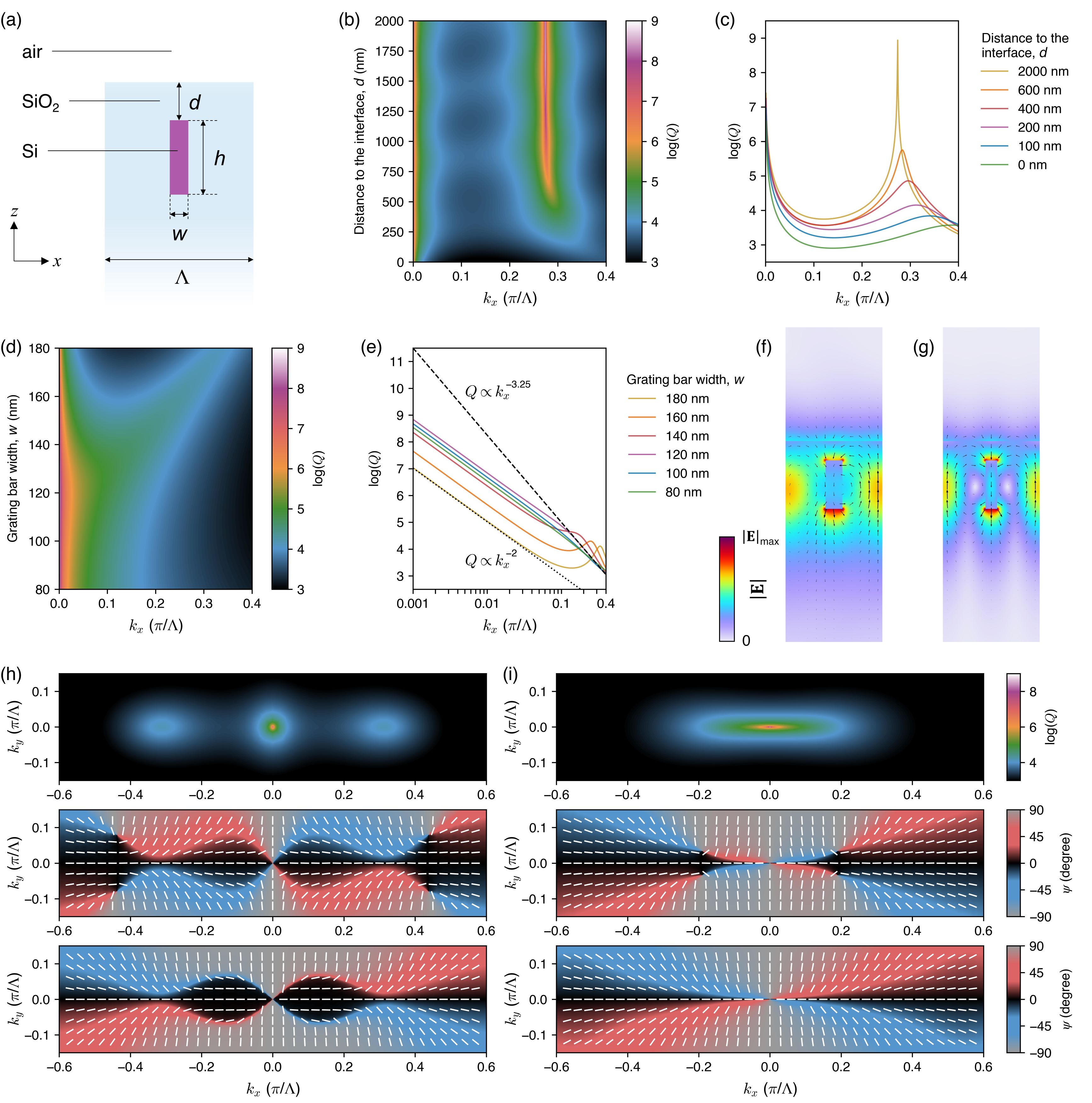}
\caption{\label{fig:3} Effects of breaking the up-down mirror symmetry on the accidental and merging BICs. (a) Schematic illustration of the unit cell of the grating, with a SiO$_2$-air interface located at a distance $d$ from the grating. (b) The $Q$ factor as a function of $k_x$ and $d$, with $w$ = 175 nm and $h$ = 500 nm. (c) Plot of the $Q$ factor versus $k_x$ for several values of $d$ [horizontal cross sections through (b)]. (d) The $Q$ factor as a function of $k_x$ and $w$, with $d$ = 200 nm and $h$ = 500 nm. (e) Log-log plot of the $Q$ factor versus $k_x$ for several values of $w$ [horizontal cross sections through (d)]. The scaling laws of $Q\propto1/k^{3.25}$ and $Q\propto1/k^2$ are shown by the dashed and dotted black lines, respectively. (f) and (g) Electric field distributions (color for $\left|\mathbf{E}\right|$, arrows for instantaneous $\mathbf{E}$) of the accidental and merging (quasi-)BICs, respectively. (h) and (i) Momentum-space distributions of the $Q$ factor (top plots) and the orientation of the long axis of the polarization ellipse [Eq.(\ref{eq:psi})] in the upward (middle plots) and downward (bottom plots) radiation before and after BIC merging, respectively.}
\end{figure*}

Optical near-fields of merging BICs in the above-considered structures are buried deep in SiO$_2$, and as such, cannot be used. To make the BIC fields accessible for applications, the grating should be suspended in air, which is difficult to achieve in practice. Most of the photonic platforms involve structures deposited on a substrate~\cite{schulz24}. Even if the use of a substrate can be avoided, the integration of a functional material or a device with the structure supporting the BICs would be very difficult without breaking its $\sigma_z$ symmetry. Breaking this symmetry, e.g., by an interface with air illustrated in Figs. \ref{fig:1}(b) and \ref{fig:3}(a) will destroy the accidental BICs, transforming them into finite-$Q$ quasi-BICs. We track this transition in Figs. \ref{fig:3}(b) and \ref{fig:3}(c) by gradually decreasing the distance $d$ between the interface and the grating (keeping unchanged $w$ = 175 nm and $h$ = 500 nm). The interface introduces a noticeable radiation loss already at $d$ = 600 nm (decreasing the $Q$ factor below 10$^6$), whereas at $d$ = 0 the $Q$ factor of the accidental BICs is reduced nearly down to that of the background photonic band (10$^3$). On the other hand, the symmetry-protected BIC is not affected by the $\sigma_z$-symmetry breaking, retaining its diverging $Q$ factor. 

Fixing $d$ at 200 nm, we end up with a $Q$ factor on the order of $10^4$. In Figs. \ref{fig:3}(d) and \ref{fig:3}(e), we show our attempt to merge these quasi-BICs with the symmetry-protected BIC by tuning $w$, just like in the previous case [see Figs. \ref{fig:2}(c) and \ref{fig:2}(d)]. We find that the $Q$ factor at $k_x$ = 0.001 $\pi/\Lambda$ can be increased by up to two orders of magnitude at $w$ = 120 nm, whereas the scaling law is improved only to slightly exceed $1/k_x^{3}$, and only at $k_x>$ 0.1 $\pi/\Lambda$. 

Figures \ref{fig:3}(f) and (g) show the near-fields of the accidental and merging (quasi-)BICs in such a system. In both cases, the evanescent fields of the modes extend above the interface, allowing them to be physically accessed from the top. However, this comes at the cost of the reduced $Q$ factor and deteriorated topological robustness of the BICs.

The far-field polarization maps in Figs. \ref{fig:3}(h) and \ref{fig:3}(i) reveal the reason of the reduced performance. Breaking the $\sigma_z$ symmetry splits the polarization vortices of the accidental BICs into half-vortices (corresponding to circularly-polarized states of a half-integer topological charge) moved towards positive and negative $k_y$. These states are associated with outgoing radiation, rendering the $Q$ factors finite. Moreover, their momentum-space trajectories are different in the upward and downward radiation (see the middle and bottom plots, respectively). For example, in the upward radiation in Fig. \ref{fig:3}(h), the half-vortices move away from the $\Gamma$-point, while in the downward radiation, they merge with the symmetry-protected BIC, flipping its topological charge from $-$1 to $+$1. Large difference in the momentum-space positions of the half-vortices in the upward and downward radiation makes it impossible to merge them with the symmetry-protected BIC vortex simultaneously in both radiation directions, as can be seen in Fig. \ref{fig:3}(i). Therefore, the overall $Q$ factor can only be slightly increased by merging, due to an improved suppression of the outgoing radiation in one direction but not in the other direction.

\begin{figure*}[hbt!]
\centering\includegraphics[width=14.788cm]{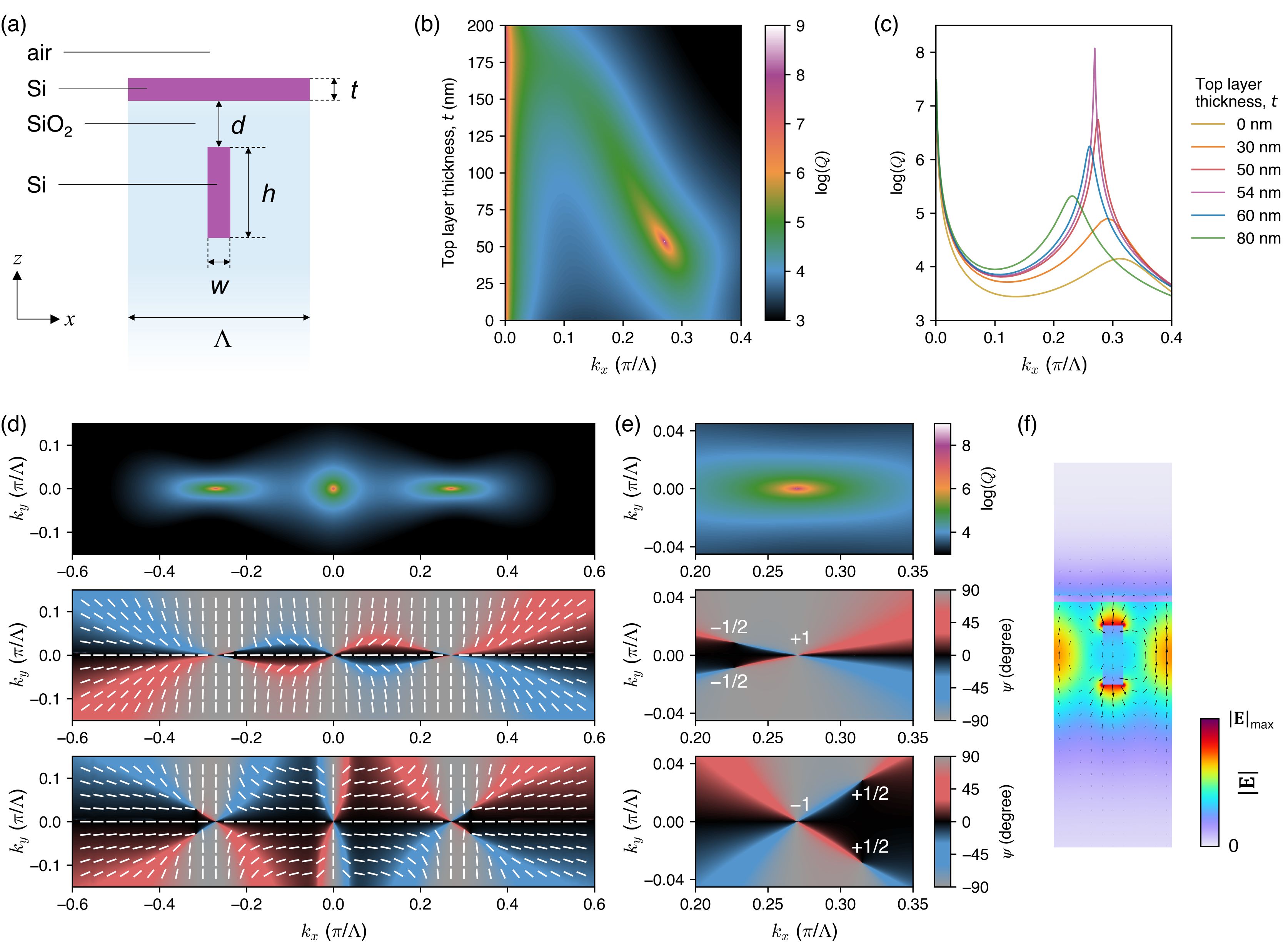}
\caption{\label{fig:4} Restoring the diverging $Q$ factor of the accidental BICs. (a) Schematic illustration of the unit cell of the grating with a top layer of thickness $t$ to compensate for the symmetry breaking. (b) The $Q$ factor as a function of $k_x$ and $t$, with $w$ = 175 nm, $h$ = 500 nm, and $d$ = 200 nm. (c) Plot of the $Q$ factor versus $k_x$ for several values of $t$ [horizontal cross sections through (b)]. (d) Momentum-space distribution of the $Q$ factor (top plot) and the orientation of the long axis of the polarization ellipse [Eq.(\ref{eq:psi})] in the upward (middle plot) and downward (bottom plot) radiation. (e) Close-ups of the momentum-space region in (d) showing the restored accidental BIC and the accompanying half-vortices. (f) Electric field distribution (color for $\left|\mathbf{E}\right|$, arrows for instantaneous $\mathbf{E}$) of the restored accidental BIC. }
\end{figure*}

\begin{figure}[hbt!]
\centering\includegraphics[width=7.59266cm]{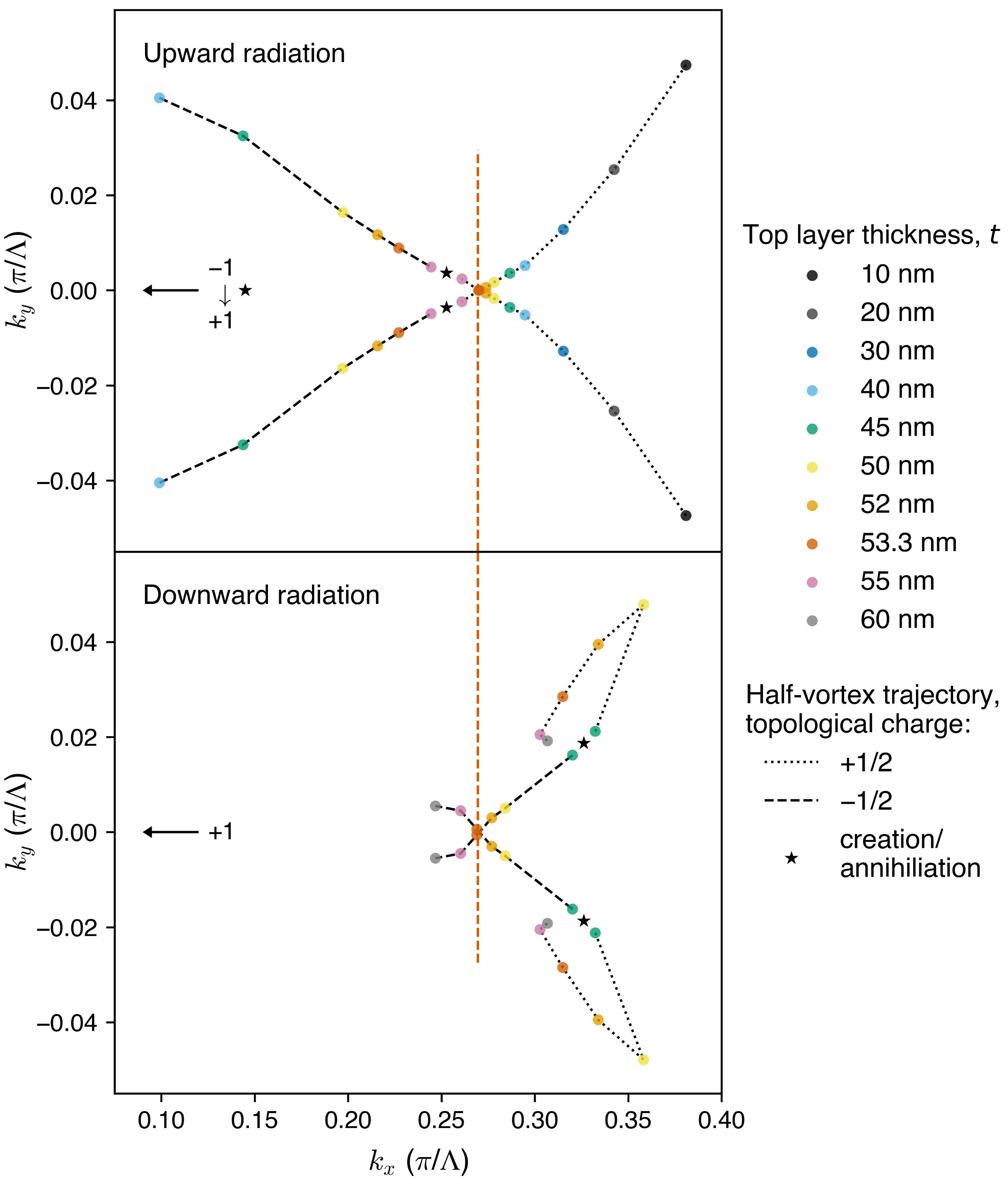}
\caption{\label{fig:5} Momentum-space trajectories of the half-vortices of charge $+$1/2 (black dotted lines) and $-$1/2 (black dashed lines) upon increasing thickness $t$ of the top layer (indicated by the colors), for the same grating geometry as in Fig. \ref{fig:4} with $w$ = 175 nm, $h$ = 500 nm, and $d$ = 200 nm. In the upward radiation, two pairs of oppositely charged half-vortices annihilate each other at $t>$ 55 nm, whereas in the downward radiation, two pairs of oppositely charged half-vortices are created at $t>$ 40 nm. The arrows point at the symmetry-protected BIC at the $\Gamma$-point (which is outside the plotted $k_x$ range). In the upward radiation, the charge of the symmetry-protected BIC is changed from $-$1 to $+$1 as a result of giving birth to four half-vortices at $t>$ 30 nm (two moving towards positive $k_x$ and two towards negative $k_x$). The in-plane symmetry of the grating ensures that the half-vortices intersect at $k_y$ = 0 simultaneously in both the upward and downward radiation directions (at $t$ = 53.3 nm, indicated by the vertical red dashed line), restoring the full vortex of the accidental BIC with a diverging $Q$ factor.}
\end{figure}

\subsection{Recovery of the accidental BICs}

Recovery of BICs in various symmetry-broken systems has been proposed and demonstrated in several previous works~\cite{yin20,yuan21,xu24}. For example, in Ref.~\cite{yin20}, the authors managed to restore accidental BICs in asymmetric periodic gratings with oblique walls by overlapping half-vortices in only one radiation direction (upward or downward), giving rise to a ``unidirectional BIC''. In Ref.~\cite{yuan21}, BICs were found to exist in 1D periodic gratings with broken in-plane symmetry for specific sets of two geometrical parameters along continuous contours in the parameter space. In Ref.~\cite{xu24}, the authors proposed a method based on a tapered geometry to restore the $Q$-factors of quasi-BICs deteriorated by the presence of a substrate. 

However, all of the above examples involve complex geometry modifications that are very challenging to implement in practice. In contrast, we propose to compensate for the broken $\sigma_z$ symmetry by adding a thin layer of a high-refractive-index material, as shown in Figs. \ref{fig:1}(c) and \ref{fig:4}(a). Intuitively, such a layer is intended to effectively balance the dielectric environment on both sides of the grating~\cite{bai24_}. Here, we consider the layer to be made of the same material (Si) as the grating, but one could choose any other high-index material, e.g., a functional material. Figures \ref{fig:4}(b) and \ref{fig:4}(c) show that the diverging $Q$ factor of the accidental BIC is restored if the Si layer thickness is $t$ $\approx$ 54 nm. Without the layer ($t$ = 0), the grating corresponds to that presented in Fig \ref{fig:3} (with $w$ = 175 nm, $h$ = 500 nm, $d$ = 200 nm), hosting the quasi-BIC with a $Q$ factor limited to $10^4$. It turns out that adding the layer of almost any $t$ provides a noticeable improvement of the $Q$ factor. For example, if $t$ is off by $\pm$ 25 nm from the optimum, the $Q$ factor can still reach $10^5$, while if $t$ is off by $\pm$ 5 nm, one can achieve $Q$ on the order of $10^6$. In addition, increasing the layer thickness shifts the accidental (quasi-)BIC along the $k_x$-axis, up to its merging with the symmetry-protected BIC at $t$ close to 200 nm.

Figures \ref{fig:4}(d) and \ref{fig:4}(e) show that the radiation pattern of the restored accidental BIC is actually not trivial. For example, the BIC has an opposite topological charge in the upward and downward radiation. It is also accompanied by a pair of half-vortices that neutralizes the overall topological charge in its vicinity in the momentum space. At the same time, the topological charge of the symmetry-protected BIC is flipped to $+$1 in both the upward and downward directions, meaning that it must have merged with other topologically-charged features. Despite the complexity of the far-field polarization patterns, the near-field [Fig. \ref{fig:4}(f)] appears to be similar to that of the accidental BIC in the $\sigma_z$-symmetric grating [Fig. \ref{fig:2}(e)]. It is also noticeably more up-down symmetric compared to the symmetry-broken case [Fig. \ref{fig:3}(f)].

To explain the observed BIC recovery, we investigate into the momentum-space evolution of the half-vortices upon gradually increasing $t$, as shown in Fig. \ref{fig:5}. In the upward radiation, two half-vortices of charge $+$1/2 move towards the line $k_y$ = 0, intersecting with it and restoring the full $+$1 vortex at $t$ = 53.3 nm. Further increase of $t$ separates the half-vortices again, moving them towards another pair of half-vortices of charge $-$1/2. These half-vortices have splitted from the symmetry-protected BIC of charge $-1$, leaving behind charge $+$1 (a second identical pair of half-vortices has moved towards negative $k_x$). Finally, all the oppositely-charged half-vortices annihilate each other at $t>$ 55 nm. 

In contrast, in the downward radiation, there are no half-vortices at small $t$ (at least in the studied momentum-space region). Only when $t$ is increased above 40 nm, pairs of oppositely-charged half-vortices are created. Those with charge $-$1/2 move towards the line $k_y$ = 0 and intersect with it at $t$ = 53.3 nm, forming a full vortex of charge $-$1. The intersection coincides with that observed in the upward radiation, not only in $t$, but also in $k_x$, which is why the diverging $Q$ factor of the accidental BIC is fully restored. This coincidence is enforced by the in-plane $C_2^zT$ symmetry of the grating structure. To restore a full vortex only in one radiation direction, this symmetry would have to be broken~\cite{yin20}, but in our case it is preserved.

\begin{figure*}[hbt!]
\centering\includegraphics[width=13.251cm]{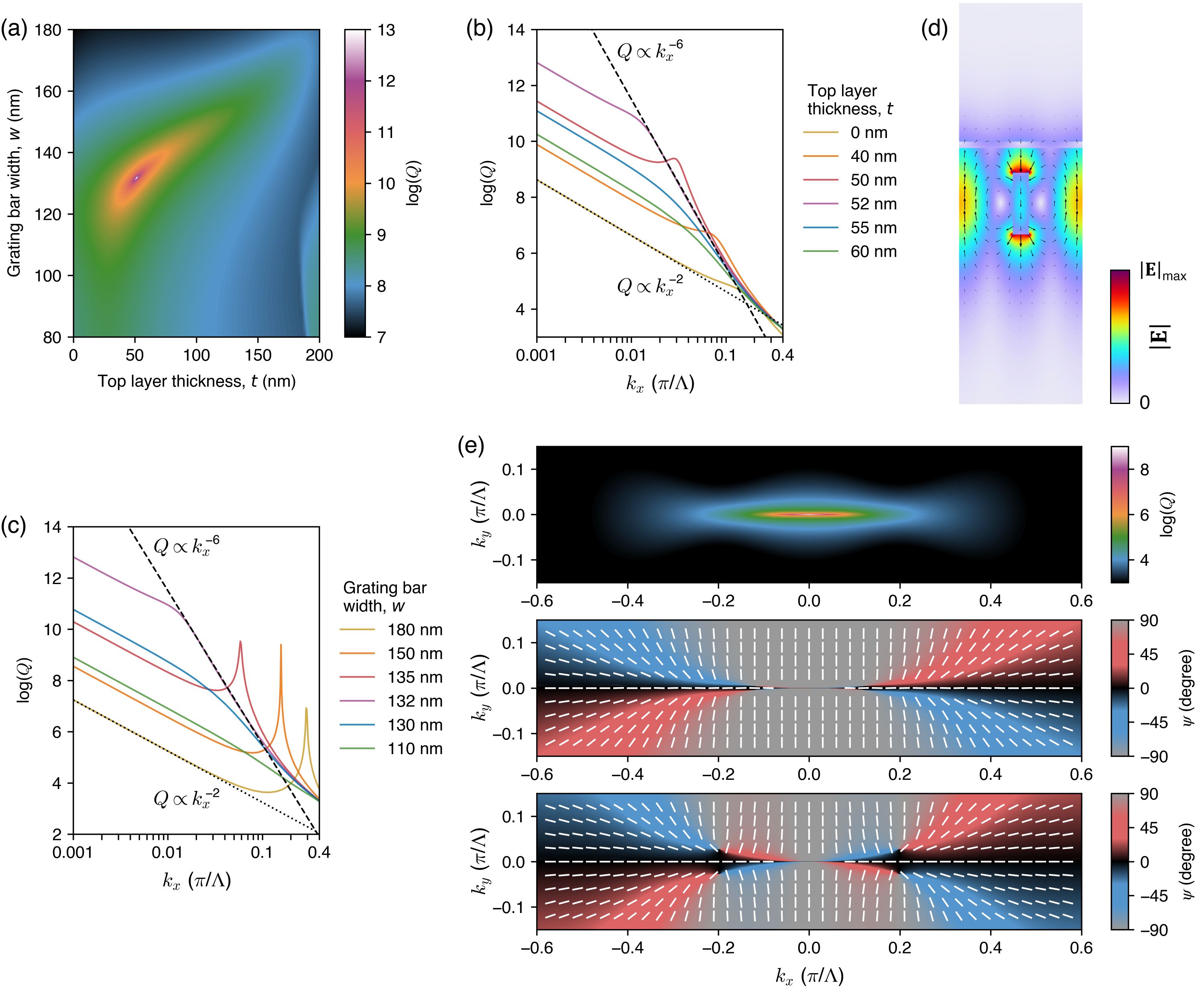}
\caption{\label{fig:6}Restoring the extreme $Q$ factor of the merged BIC. The grating geometry is the same as in Fig. \ref{fig:4}(a). (a) The $Q$ factor at $k_x$ = 0.001 $\pi/\Lambda$ as a function of $t$ and $w$, with $h$ = 500 nm and $d$ = 200 nm. (b) and (c) Log-log plots of the $Q$ factor versus $k_x$ for several values of $t$ and $w$, respectively. The scaling laws of $Q\propto1/k^{6}$ and $Q\propto1/k^2$ are shown by the dashed and dotted black lines, respectively. (d) Electric field distribution (color for $\left|\mathbf{E}\right|$, arrows for instantaneous $\mathbf{E}$) of the restored merged BIC. (e) Momentum-space distribution of the $Q$ factor (top plot) and the orientation of the long axis of the polarization ellipse [Eq.(\ref{eq:psi})] in the upward (middle plot) and downward (bottom plot) radiation. }
\end{figure*}

\subsection{Recovery of the merged BIC}

Complete recovery of the accidental BICs means that it should be possible to restore also the extreme $Q$-factor boost of the merged BIC. However, as opposed to the $\sigma_z$-symmetric grating considered at the beginning of this paper, the restored accidental BIC cannot be simply moved along $k_x$ by tuning $w$, as this would alter the optimal value of $t$. To locate the new merging point, we repeat the same procedure as in Fig. \ref{fig:2}(g), looking at the values of the $Q$ factor at a fixed $k_x$ = 0.001 $\pi/\Lambda$. Figures \ref{fig:6}(a), \ref{fig:6}(b), and \ref{fig:6}(c) present these values as a function of $t$ and $w$, revealing the merging point at $t\approx$ 52 nm and $w \approx$ 132 nm. At this point, the $Q$-factor scaling is improved to $1/k_x^6$, and the merging-induced boost of the $Q$ factor is close to that observed in the $\sigma_z$-symmetric gratings (Fig. \ref{fig:2}), with the values of the $Q$ factor approaching $10^{13}$ at $k_x$ = 0.001 $\pi/\Lambda$, i.e., six orders of magnitude higher than without merging. 

The near-field of the merged BIC [Fig. \ref{fig:6}(d)] is very similar to that in the $\sigma_z$-symmetric grating [Fig. \ref{fig:2}(f)] and clearly more symmetric than in the grating with broken $\sigma_z$ symmetry [Fig. \ref{fig:3}(g)]. On the other hand, the far-field pattern still lacks the up-down symmetry [see Fig. \ref{fig:6}(e)]. While the upward radiation shows a fully merged BIC of charge $+$1, the downward radiation features a BIC of charge $-1$ surrounded by half-vortices. The presence of additional half-vortices is not surprising, taking into account their different locations around the restored accidental BIC [Fig. \ref{fig:4}(e)] as well as the fact that they can be created and annihilated independently in the upward and downward radiation (Fig. \ref{fig:5}). The process of half-vortex creation and annihilation has recently been shown to originate from an interplay between BICs and Dirac points on degenerate photonic bands~\cite{zhang24}. We do not discuss the exact mechanism in our case, leaving it as a topic for future investigations.

\subsection{Strongly broken symmetry ($d$ = 0)}

\begin{figure*}[hbt!]
\centering\includegraphics[width=17.8cm]{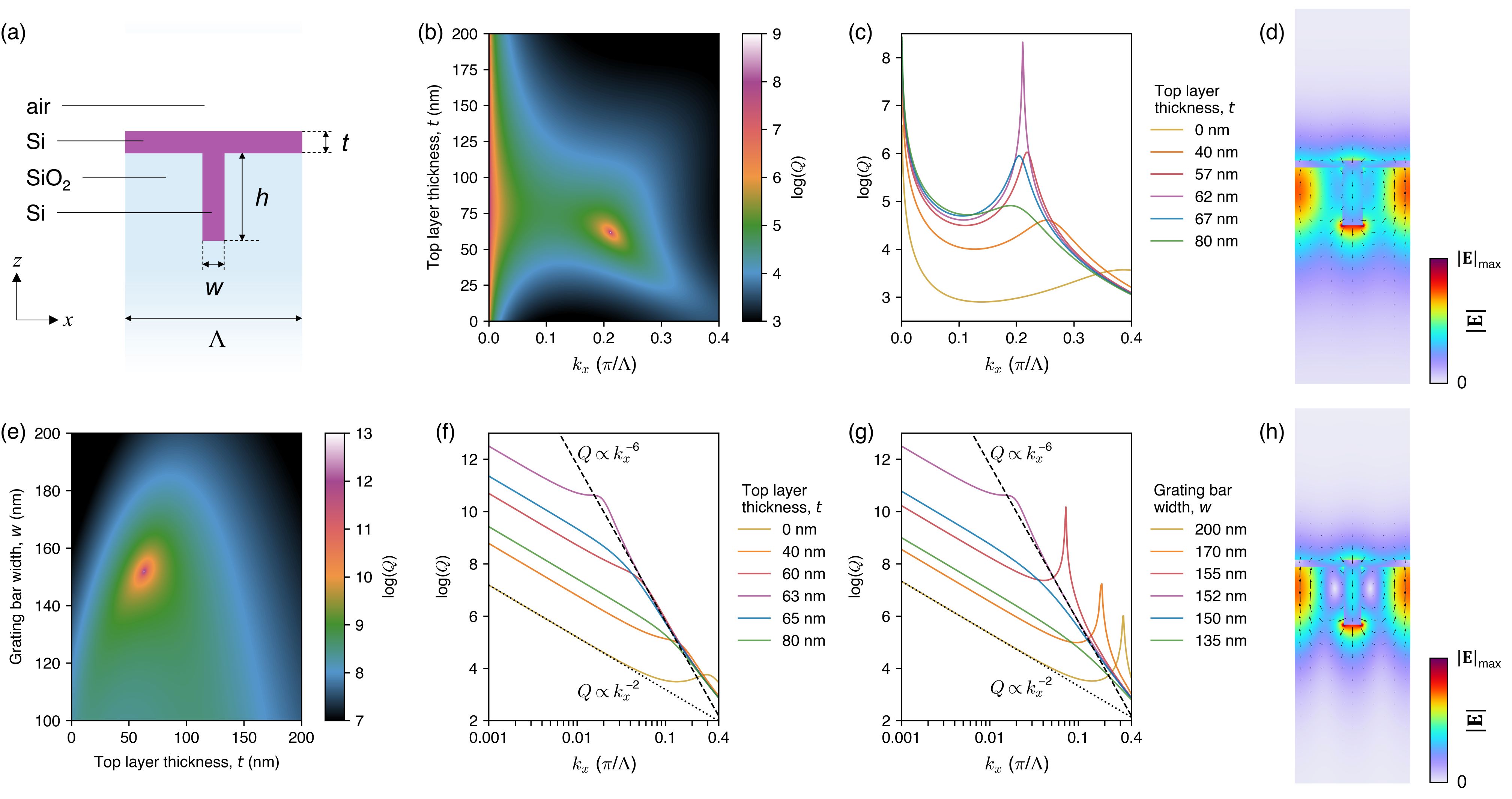}
\caption{\label{fig:7} Restoring the accidental and merging BICs in a grating with strongly broken up-down mirror symmetry. (a) Schematic illustration of the unit cell of the grating, which is equivalent to the geometry in Fig. \ref{fig:4}(a) with $d$ = 0. (b) The $Q$ factor as a function of $k_x$ and $t$, with $w$ = 175 nm and $h$ = 500 nm. (c) Plot of the $Q$ factor versus $k_x$ for several values of $t$ [horizontal cross sections through (b)]. (d) Electric field distribution (color for $\left|\mathbf{E}\right|$, arrows for instantaneous $\mathbf{E}$) for the restored accidental BIC. (e) The $Q$ factor at $k_x$ = 0.001 $\pi/\Lambda$ as a function of $t$ and $w$, with $h$ = 500 nm. (f) and (g) Log-log plots of the $Q$ factor versus $k_x$ for several values of $t$ and $w$, respectively. The scaling laws of $Q\propto1/k^{6}$ and $Q\propto1/k^2$ are shown by the dashed and dotted black lines, respectively. (h) Electric field distribution (color for $\left|\mathbf{E}\right|$, arrows for instantaneous $\mathbf{E}$) of the restored merged BIC.}
\end{figure*}

The electric field distributions presented in Figs. \ref{fig:4}(f) and \ref{fig:6}(d) show that the compensation for the broken symmetry makes the restored accidental and merging BICs much more confined inside the grating. In particular, the amplitude of the evanescent field extending into the air is significantly diminished, compared to that of the BICs without the compensating layer [see Figs. \ref{fig:3}(f) and \ref{fig:3}(g)]. Accessing the near-fields of BICs is essential for many practical applications, e.g., to enhance light-matter interactions in sensing. Therefore, we consider a geometry shown in Fig. \ref{fig:7}(a), which leaves no gap between the grating and the top layer ($d$ = 0), pushing one of the hot spots outside the grating. Figures \ref{fig:7}(b) and \ref{fig:7}(c) [analogous to \ref{fig:4}(b) and \ref{fig:4}(c)] demonstrate that the diverging $Q$ factor of the accidental BIC can be fully restored even in such an extreme scenario, while also giving rise to a high evanescent-field amplitude on the top surface of the structure [Fig. \ref{fig:7}(d)]. 

Figures \ref{fig:7}(e), \ref{fig:7}(f), and \ref{fig:7}(g) [analogous to \ref{fig:6}(a), \ref{fig:6}(b), \ref{fig:6}(c)] show that the merged BIC can also be fully restored in this geometry. Upon compensation for the broken $\sigma_z$ symmetry with perfectly adjusted parameters ($w$ = 152 nm, $t$ = 63 nm), BIC merging brings the $Q$-factor scaling up to $1/k_x^6$ and provides the $Q$-factor boost of almost six orders of magnitude, similar to that in a fully $\sigma_z$-symmetric grating presented in Fig. \ref{fig:2}. At the same time, the evanescent field of the merged BIC becomes physically accessible from the top, as can be seen in Fig. \ref{fig:7}(h), providing a higher relative field amplitude than in Fig. \ref{fig:6}(d). 

Figures \ref{fig:7}(b), \ref{fig:7}(c), \ref{fig:7}(e), \ref{fig:7}(f) and \ref{fig:7}(g) reveal the tolerance of the restored BICs to the geometry distortions. For example, Fig. \ref{fig:7}(c) shows that the $Q$ factor of the restored accidental BIC exceeds 10$^6$ if $t$ is off by $\pm$ 5 nm, and it remains close to 10$^5$ if $t$ is off by $\pm$ 20 nm. Furthermore, the $Q$-factor boost by BIC merging remains very high across a wide range of parameters. For example, if the values of $t$ or $w$ are off by $\pm$ 20 nm, the $Q$ factor at $k_x$ = 0.001 $\pi/\Lambda$ is still improved by approximately two orders of magnitude. This means that the proposed method enables one to achieve exceptionally high $Q$ factors in experimentally realizable structures.

\section{Conclusions}

In this work, we have shown that diverging $Q$ factors of accidental BICs and topological robustness of merging BICs can be obtained in structures with broken up-down mirror symmetry ($\sigma_z$) with the help of a simple asymmetry compensation method. As an example, we have considered a 1D periodic dielectric grating, in which the $\sigma_z$ symmetry is broken by a one-sided interface with air. This transforms the accidental BICs into quasi-BICs and splits their polarization vortices into pairs of circularly polarized half-vortices. Although these quasi-BICs can be tuned to merge with the symmetry-protected BIC at the $\Gamma$-point, the resulting merged BIC shows a limited $Q$ factor compared to that in a $\sigma_z$-symmetric grating. By adding a layer of a high-index material on one side of the grating, we restore the integer-charge vortex of the accidental BIC. The underlying mechanism is based on the in-plane $C_2^zT$ symmetry-protected coincidence of intersections between the momentum-space trajectories of half-vortices in both upward and downward radiation. This approach allows us to fully restore also the merging-assisted $Q$-factor boost, including the improved $Q$-factor scaling law of $k_x^{-6}$. Finally, by considering a strongly asymmetric geometry, we demonstrate that the restored accidental and merging BICs offer physical access to their near-fields, which is important for practical applications based on enhanced light-matter interactions, e.g., for optical sensing, lasing, and nonlinear optical effects. 

The inherent robustness of the merging BICs makes our design highly practical. For instance, every realistic photonic structure has a finite size, which splits the BICs into many copies distributed over the momentum space. The role of the merged BIC is to boost the Q factor of all these modes simultaneously. This happens with high efficiency even before the complete merging is achieved~\cite{jin19}. Similar observations have been made in the case of a BIC laser, in which the optimal suppression of radiation loss occurs before the complete merging~\cite{hwang21}. Moreover, in practice, surface roughness and other fabrication imperfections, as well as finite absorption by the materials can easily push the achievable $Q$ factors down to $\sim$ 10$^6$ \cite{jin19,chen22}, regardless of the exact asymmetry compensation and exactly reached merging point. Due to these additional losses, the practical tolerance to the geometrical parameters is actually even higher than the tolerances discussed in this paper. 

The proposed mechanism of the BIC recovery is not limited to 1D gratings and can readily be extended to 2D photonic crystal slabs and metasurfaces, enabling ultrahigh-$Q$ accidental and merging BICs to be observed in structures without the up-down mirror symmetry. This opens the possibility to create a variety of surface-mounted structures that support topologically robust BICs with accessible near-fields. Furthermore, the top layer that compensates for the asymmetry can be made of any material that has a refractive index higher than that of the substrate. This could be a functional material, e.g., a gain medium, a nonlinear medium, or an analyte. Therefore, we believe our findings will unlock the true potential of BICs for diverse light-based technologies. Our findings also suggest that certain symmetry-protected topological properties can be achieved even if the essential symmetry is broken, which could potentially relax critical symmetry constraints in many systems.

\section*{Acknowledgment}

The authors acknowledge the support of the Academy of Finland (Grants No. 347449 and 353758). For computational resources, the authors acknowledge the Aalto University School of Science “Science-IT” project and CSC – IT Center for Science, Finland.

\section*{Data availability}

Data underlying the results presented in this paper and their replication instructions are openly available at \href{https://doi.org/10.23729/d98b1b49-c338-4ea5-9007-ebd712fa6cca}{https://doi.org/10.23729/d98b1b49-c338-4ea5-9007-ebd712fa6cca}, Ref. \cite{10.23729/d98b1b49-c338-4ea5-9007-ebd712fa6cca}.

\bibliography{lib}

\end{document}